\documentclass[a4paper,aps,showpacs,floatfix,superscriptaddress,prl,preprint]{revtex4}
\usepackage{amsmath,amssymb,eucal,graphicx}

\begin{document}

\today

\title{
Master Stability Functions for Coupled Near-Identical Dynamical Systems
}

\author{Jie Sun}
\email{sunj@clarkson.edu}
\author{Erik M. Bollt}
\email{bolltem@clarkson.edu}
\author{Takashi Nishikawa}
\email{tnishika@clarkson.edu}
\affiliation{Department of Mathematics \& Computer Science, Clarkson University, Potsdam, NY 13699-5815}

\begin{abstract}
We derive a master stability function (MSF) for synchronization in networks of coupled dynamical systems with small but arbitrary parametric variations.
Analogous to the MSF for identical systems, our generalized MSF simultaneously solves the linear stability problem for near-synchronous states (NSS) for all possible connectivity structures.
We also derive a general sufficient condition for stable near-synchronization and show that the synchronization error scales linearly with the magnitude of parameter variations.
Our analysis underlines significant roles played by the Laplacian eigenvectors in the study of network synchronization of near-identical systems.
\end{abstract}

\pacs{05.45.Xt, 89.75.Hc
}
\maketitle

{\it Introduction ---}
Synchronization in its various forms has been a highly popular and exciting developing topic in the recent literature on chaotic oscillators \cite{PECORA_PRL90,Pikovsky:2001pf}.  Applications have ranged widely from biology \cite{STROGATZ_SCIAM93, STROGATZ_BOOK03} to mathematical epidemiology \cite{He_BIOLSCI03}, and modelling animal gaits \cite{BUONO_JMATHBIOL01} to engineering of communications devices \cite{PECORA_PRL90,CUOMO_PRL93}, including many developments in complex networks 
(see e.g., Refs. \cite{NISHIKAWA_PRL03,RESTREPO_PRE04,Gomez-Gardenes:2006fq,Lu:2006fk,Zhou:2006kb} and a review \cite{BOCCALETTI_PHYSREP02}).  
However, the preponderance of the work has focused on identical synchronization since it is in this situation whereby a complete analysis can be carried forward by the master stability formalism developed in the seminal work \cite{PECORA_PRL98}.  While other forms of synchronization are discussed in the literature, 
of particular interest here is  {\it nearly-synchronous state} behavior of the systems that are slightly detuned from identical synchronization, which may or may not be associated with an invariant manifold~\cite{BOCCALETTI_PRE01} normally required to describe generalized synchrony~\cite{RULKOV_PRE95, ABAR_PRE95, PECORA_PRE95}.

In this letter we consider a coupled dynamical system consisting of $N$ units coupled through some underlying network. 
The equations of motion reads:
\begin{equation}\label{origdyn}
	\dot{w_i} = f(w_i,\mu_i) - g\sum_{j=1}^{N}{L_{ij}H(w_j)}, \quad i = 1,2,\dots, N,
\end{equation}
where
$f:\Re^{m\times p}\rightarrow \Re^{m}$ is the parameterized dynamics of an isolated unit;
$w_i\in \Re^{m}$ is the dynamical variable 
for the
$i$th unit; $\mu_i\in \Re^{p}$ is the corresponding parameter;
$L\in \Re^{N\times N}$ is the graph Laplacian
	\footnote{We only deal with graph Laplacians that are diagonalizable for the reason of clarity. Non-diagonalizable 	Laplacians can be treated by techniques proposed in \cite{NISHIKAWA_PHYSICAD06}.};
$H:\Re^{m}\rightarrow\Re^{m}$ is a uniform coupling 
function;
and $g\in \Re$ is the uniform coupling strength (usually $>0$ for diffusive coupling).

Note that we can represent the whole system conveniently by using Kronecker product representation:
\begin{eqnarray}
	\mbox{{\boldmath $\dot w$}} = \mbox{{\boldmath $f(w,\mu)$}} - g\cdot L\otimes H(\mbox{{\boldmath $w$}}),
\end{eqnarray}
where {\boldmath $w$}$=[w_1^T,w_2^T,...w_N^T]^T$ is a column vector of all the dynamic variables, and likewise for {\boldmath $\mu$} and {\boldmath $f$}; and $\otimes$ is the usual Kronecker product (or direct product)~\cite{BERBERIAN_BOOK}.

System~(\ref{origdyn}) has been studied mostly in the case 
in which the parameter $\mu_i$ is the same for each individual oscillator,
often resulting
in {\it identical synchronization} where $\max_{i,j}{||w_i(t)-w_j(t)||}\rightarrow 0$ as $t\rightarrow\infty$. The stability of 
such states
 can be analyzed by master stability functions (MSF) \cite{PECORA_PRL98}.

However, a noiseless system with exactly the same parameters is impossible 
in practice. It is known that parameter mismatch 
among the individual 
oscillators can cause bursts due to the instability of typical periodic orbits embedded in the synchronized chaotic attractor \cite{RESTREPO_PRE04}; even within a stable region where no bubbling will occur, the states of different units will still not approach exactly 
the same function of time, 
but instead come close to each other within a neighborhood of the identical synchronization 
state~\cite{RESTREPO_PRE04}. 
This phenomena was first reported in \cite{PECORA_PRL90} for two coupled Lorenz oscillators, where the variations of individual units 
from 
the identical synchronization manifold was found to scale linearly with respect to the magnitude of parameter mismatch when the mismatch is small.
In~\cite{RESTREPO_PRE04}, a variational equation analogous to our Eq.~(\ref{vareq}) was used to study the progressive loss of synchronization stability due to bursting, which is also a relevant and interesting phenomenon.
In this letter we develop an extended master stability framework for systems with near-identical parameters and derive stability conditions for stable near-synchronization.

{\it Near-Synchronous State (NSS)---}
Assume that the parameters $\mu_i$ in Eq.~(\ref{origdyn}) are close to each other and do not change with time. Let the {\it average parameter} be 
	$\bar{\mu} \equiv \frac{1}{N}\sum_{i=1}^{N}{\mu_i}$
and
the {\it parameter mismatch} be
	$\delta\mu_i \equiv \mu_i - \bar{\mu}$.
With appropriate choices of coupling strength $g$ and network structure $L$, the system can have a {\it near synchronous state} (NSS) in which
	$\max_{i,j}{||w_i(t)-w_j(t)||} \leq c \mbox{ }\mbox{ as }\mbox{ t} \rightarrow\infty$
for some small constant $c\geq 0$. 
When the system undergoes such near-synchronization, 
the trajectories of individual units are well approximated by 
the {\it average trajectory}
	$\bar{w} \equiv \frac{1}{N}\sum_{i=1}^{N}{w_i}$,
which is governed by
\begin{equation}
	\dot{\bar{w}} = \frac{1}{N}\sum_{i=1}^{N}{\dot{w_i}}
					= \frac{1}{N}\sum_{i=1}^{N}{f(w_i,\mu_i)} - g\sum_{j=1}^{N}{d_{j}H(w_j)},
\end{equation}
where we have defined
	$d_j \equiv \frac{1}{N}\sum_{i=1}^{N}{L_{ij}}$
\footnote{For undirected graphs, we have 
$d_j = 0$ 
for all $j$ since $L$ is symmetric with each row sum equalling zero.}. 
With this equation, 
we can discuss dynamics of the bulk, or coarse scale behavior.

{\it Inhomogeneity in Variational Equations ---}
Define the variation on each individual unit to be
	$\eta_{i} \equiv w_{i} - \bar{w}$
for $i=1,2,...,N$. The
variational equations is then
\begin{eqnarray}	
	\dot{\eta_i} &=& \Big[f(\bar{w}+\eta_i,\bar{\mu}+\delta\mu_i)-\frac{1}{N}\sum_{j=1}^{N}{f(\bar{w}+\eta_j,\bar{\mu}+\delta\mu_j)} \Big] \nonumber\\ 
	&&- g\sum_{j=1}^{N}{(L_{ij}-d_j)H(\bar{w}+\eta_j)}.
\end{eqnarray}
Assuming that the variations $\eta_i$ and the parameter mismatch $\delta\mu_i$ are small, we expand around $\bar{w}$ and $\bar{\mu}$ to obtain 
\begin{eqnarray}
	\dot{\eta_i} &=& D_{w}f(\bar{w},\bar{\mu}) \eta_i  - g\sum_{j=1}^{N}{(L_{ij}-d_j)DH(\bar{w}) \eta_j} \nonumber\\
	&& + D_{\mu}f(\bar{w},\bar{\mu}) \delta\mu_i.
\end{eqnarray}
We have used
	$\sum_{j=1}^{N}{\eta_j} \equiv \sum_{j=1}^{N}{w_j} - N\cdot\bar{w} = 0$,
	$\sum_{j=1}^{N}{\delta\mu_j} \equiv \sum_{j=1}^{N}{\mu_j} - N\cdot\bar{\mu} = 0$, and 
	$\sum_{j=1}^{N}{d_j} \equiv \sum_{i,j}{L_{ij}} = 0$
in the derivation.
Putting all the $\eta_i$ and $\delta\mu_i$ in column vectors {\boldmath$\eta$} and {\boldmath$\delta\mu$}, respectively,
and omitting the arguments $(\bar{w},\bar\mu)$
for simplicity, we obtain the 
{\it variational equation for the NSS}:
\begin{eqnarray}\label{vareq}
	\mbox{{\boldmath $\dot\eta$}} = \Big[ I_N\otimes D_{w}f - g\cdot G \otimes DH \Big]\mbox{{\boldmath $\eta$}} + 
	\Big[ I_N \otimes D_{\mu}f \Big]\mbox{{\boldmath $\delta\mu$}},
\end{eqnarray}
where the {\it modified graph Laplacian} $G$ is defined by 
$G\equiv L - [1,1,...,1]^{T} \cdot [d_1,d_2,...d_N]$ 
and $I_N$ is the 
$N \times N$
identity matrix.

Interestingly, the 
eigenvalues of $G$, $\lambda_1, \lambda_2, \ldots, \lambda_N$ are exactly the same as those of $L$, and 
the vector 
$[1,1,...,1]^{T}$ is the eigenvector of both $L$ and $G$
associated with $\lambda_1 = 0$.
Furthermore, 
any other eigenvector 
$v'$
of $G$ associated with eigenvalue $\lambda$
can be obtained from the corresponding eigenvector 
$v$ 
of $L$ by 
the transformation 
	$v' = v - [\bar{v},...,\bar{v}]^T$,
which simply shifts each
component of 
$v$ 
by a constant 
$\bar{v} = \frac{1}{\lambda}\sum_{j=1}^{N}{d_j v_j}$. 
More importantly,
if there exist diagonalization forms $L = Q\Lambda Q^{-1}$ and $G = P\Lambda P^{-1}$, 
then the corresponding rows of $Q^{-1}$ and $P^{-1}$ (the {\em left} eigenvectors of $L$ and $G$, respectively) are parallel to each other, 
except for the first rows that correspond to 
$\lambda_1 = 0$. 

When all the parameters $\mu_i$ are the same, the second term in the Eq.~(\ref{vareq}) disappears, and what is left is a homogeneous ODE system 
for {\boldmath $\eta$}, which may be diagonalized to obtain an equation 
analogous to the well-known master stability equation \cite{PECORA_PRL98}, 
with the only difference that 
here we have a modified graph Laplacian $G$. Interestingly, in the case of no parameter mismatch, 
this difference would not lead to different conclusions
since the stability analysis depends on the graph structure only through the Laplacian eigenvalues,
not eigenvectors.

We now focus on the case in which, if there were no parameter mismatch, the system would undergo stable identical synchronization, i.e., the variation {\boldmath $\eta$} would go to zero asymptotically.  This situation occurs if the system represented by $f, H, L$ and $g$ are in the stable regime~\cite{PECORA_PRL98}. 
Because of the inhomogeneous part $\big[ I_N \otimes D_{\mu}f \big]${\boldmath $\delta\mu$} due to parameter mismatch, the variational system~\eqref{vareq} in general may not be asymptotically stable. 
We will show, however, that when the parameter mismatch is small, there may exist a NSS where {\boldmath $\eta$} stays close (although not equalling) to zero.
Indeed, we will show that the variational system is stable (i.e. the solution {\boldmath $\eta$} is bounded as $t\rightarrow\infty$) and the 
bound for the solution 
depends linearly on the norm of the parameter mismatch {\boldmath $\delta\mu$}.

%

{\it Extended Master Stability Equation and Function ---}
We may uncouple the variational equation by diagonalizing 
the modified graph Laplacian $G$: $G = P\Lambda P^{-1}$
\footnote{In the case of undirected graphs, we have $P^{-1}=P^{T}$ and thus rows of $P^{T}$ correspond to eigenvectors of $L$.} for some invertible matrix $P$. 
Making the change of variable 
{\boldmath$\zeta = $} $(P^{-1}\otimes I_m)${\boldmath$\eta$}, we obtain
\begin{eqnarray}\label{vareq2}
	\mbox{{\boldmath $\dot\zeta$}} = \Big[ I_N \otimes D_{w}f - g\cdot \Lambda \otimes DH \Big] \mbox{{\boldmath $\zeta$}}
	+ \Big[ P^{-1} \otimes D_{\mu}f \Big] \mbox{{\boldmath $\delta\mu$}}.
\end{eqnarray}
The homogeneous part in Eq.~(\ref{vareq2}) has block diagonal structure and we may write for each eigenmode 
$i \ge 2$
\begin{eqnarray}\label{vareq3}
	\dot{\zeta_i} = \Big[D_{w}f - g\lambda_i DH \Big]\zeta_i 
		+ D_{\mu}f \cdot \sum_{j=1}^{N}{u_{ij}\delta\mu_j}, 
\end{eqnarray}
where 
$u_{ij}$ 
is the $j$th component of the $i$th row in the matrix $P^{-1}$, i.e.,  
$u_i$ is the $i$th left eigenvector of $G$.
The vector 
$\sum_{j=1}^{N}{u_{ij} \delta\mu_j}$ 
is the weighted average of parameter mismatch vectors, 
with the weights given by the components of the left eigenvector
associated with 
$\lambda_i$.  It may also be 
thought of as an inner product of the parameter mismatch vector and the corresponding 
left
eigenvector. We comment here that 
if one uses the original graph Laplacian $L$ instead, the resulting equation would be equivalent to Eq.~(\ref{vareq3}), since the spectra of $L$ and $G$ are the same
and corresponding left eigenvectors are parallel except for those associated with $\lambda_1=0$. 


From Eq.~(\ref{vareq3}), we define an
{\it extended master stability equation} for near identical coupled dynamical systems:
\begin{eqnarray}\label{extmsf}
	\dot{\xi} = \Big[ D_{w}f - \alpha\cdot DH \Big]\xi + D_{\mu}f\cdot\psi,
\end{eqnarray}
where we have introduced two auxiliary parameters,  a (complex) scalar $\alpha$ and $\psi\in\Re^{p}$. 
Once the stability of Eq.~\eqref{extmsf} is determined as a function of $\alpha$ and $\psi$, the stability of the $i$th eigenmode can be found by simply setting $\alpha = g \lambda_i$ and 
$\psi = \sum_{j=1}^{N} u_{ij}\delta\mu_j$.
The problem is thus decomposed into two separate parts: one that depends only on the individual dynamics and the coupling function, and the other that depends only on the graph Laplacian and parameter mismatch.  Note that the latter not only depends on the spectrum of $L$ as in \cite{PECORA_PRL98}, but also on the combination of the left eigenvectors and parameter mismatch. 
Thus, we have reduced the stability analysis of the original $mN$-dimensional problem to that of $m$-dimensional problem with one additional parameter, combined with an eigenproblem.

Note that to analyze the stability of the original system using 
the master stability equation, 
we need the associated 
average trajectory $\bar{w}$, 
which can only be obtained by solving the original system, 
and is impractical for large networks.  We found, however, that 
in practice (as we will confirm in an example below)
one may instead use a trajectory $s$ of a single auxiliary {\it average unit}:
	$\dot{s} = f(s,\bar{\mu})$.
We 
conjecture that under 
suitable conditions on the system,
the trajectory $s$ of the average unit shadows the average trajectory $\bar{w}$~\footnote{The supporting analysis and results will be reported in future work.}.

The associated {\it master stability function} $\Omega(\alpha,\psi)$ 
is then defined 
to be the asymptotic value of the norm of $\xi$ as a function of $\alpha$ and $\psi$, given that $\alpha$ leads to asymptotic stable solution 
of the homogeneous part. 
In the case of symmetrically coupled networks, for which $G=L$ is symmetric, the matrix $P$ can be chosen to be orthogonal, allowing us to predict the square-sum synchronization error in the original system~(\ref{origdyn}) from $\Omega(\alpha,\psi)$:
\begin{equation}\label{err}
\sum_{i=1}^N || \eta_i (t)||^2 = \sum_{i=2}^N || \zeta_i (t)||^2 
\xrightarrow{t \to \infty} \sum_{i=2}^N \Omega(\alpha_i, \psi_i)^2,
\end{equation}
where $\alpha_i$ and $\psi_i$ correspond to the $i$th eigenmode and $||\cdot||$ denotes the Euclidean norm.

{\it Conditions for Stable Synchronization ---}
In the previous section we have derived a generic stability equation~(\ref{extmsf}) for analyzing the stability of synchronization of coupled dynamical system~(\ref{origdyn}). To analyze the stability, we now assume that the largest Lyapunov exponent of the synchronous trajectory associated with the homogeneous variational equation
\begin{equation}\label{homomsf}
	\dot{\xi} = \Big[ D_{w}f - \alpha DH \Big]\xi
\end{equation}
is negative for a given $\alpha$, 
so that without parameter mismatch the error mode corresponding to this specific $\alpha$ goes to zero exponentially. In this case, the solution $\xi^{*}$ of Eq.~(\ref{homomsf}) can be written as:
	$\xi^{*}(t) = \Phi(t,0)\xi(0)$,
where $\Phi(t,\tau)$ is the fundamental transition matrix \footnote{This transition matrix, as a function of two time variables $t$ and $\tau$, can be obtained by the {\it Peano-Baker series} as long as $D_{w}f-\alpha DH$ is continuous. See 
\cite{RUGH_BOOK} (Ch.~3, p.~40).
}, satisfying
\begin{equation}\label{bdphi}
	||\Phi(t,\tau)|| \leq \gamma e^{-\lambda(t-\tau)}
\end{equation}
for $t\geq\tau$ and some finite positive constants $\gamma$ and $\lambda$.
We should note that in the case of generalized synchrony, the loss of stability of the invariant manifold
need not proceed monotonically and uniformly in space.
It is known that parameter mismatch can cause bursting due to increasing instability of embedded transversely unstable periodic orbits which cause short-time positivity of Lyapunov exponents~\cite{RESTREPO_PRE04, KIM_PTP02}, and this can be correspondingly interpreted from Eq.~(\ref{bdphi}).  Such transition has been called bubbling bifurcation~\cite{VENK_PRL96, VENK_PRE96} due to basin riddling.

The solution to Eq.~(\ref{extmsf}) can then be expressed by
\begin{equation}\label{solinhomo}
	\xi(t) = \Phi(t,0)\xi(0) + \int_{0}^{t}{\Phi(t,\tau)b(\tau)d\tau},
\end{equation}
where $b(\tau)\equiv D_{\mu}f(s(\tau),\bar{\mu})\cdot\psi$.
%
Under the condition of Eq.~(\ref{bdphi}), 
we can show that $\xi(t)$ given by Eq.~(\ref{solinhomo}) is bounded by the following inequality:
\begin{eqnarray}
	||\xi(t)|| &\leq& ||\Phi(t,0)||\cdot ||\xi(0)||
			+ \int_{0}^{t}{||\Phi(t,\tau)||d\tau} \cdot \sup_{t}{||b(t)||} \nonumber\\
		&\leq& \gamma e^{-\lambda t}||\xi(0)||
			+ \frac{\gamma}{\lambda}(1-e^{-\lambda t})\sup_{t}{||b(t)||}	\nonumber\\
		&\rightarrow& \frac{\gamma}{\lambda}\sup_{t}{||b(t)||}
			\mbox{ }\mbox{ as }\mbox{ }t\rightarrow\infty.
\end{eqnarray}
Thus, the inhomogeneous master stability equation is stable, i.e., the solution to Eq.~(\ref{extmsf}) is bounded asymptotically as long as
$i)$ the homogeneous system is exponentially stable, or equivalently, the maximal Lyapunov exponent  is negative; and 
$ii)$ the inhomogeneous part $b(\tau)\equiv D_{\mu}f(s(\tau),\bar{\mu})\cdot\psi$ is bounded.

Eq.~(\ref{bdphi}) and Eq.~(\ref{solinhomo}) also allow us to analyze quantitatively the magnitude of asymptotic error of a near-identical 
system. 
If the magnitude of parameter mismatch is scaled by a factor $c$, keeping all other 
parameters fixed,
it follows from Eq.~(\ref{solinhomo}) that the corresponding solution will be
\begin{equation}\label{solinhomo2}
	\widetilde{\xi}(t) = \Phi(t,0)\xi(0) + c\int_{0}^{t}{\Phi(t,\tau)b(\tau)d\tau},
\end{equation}
where $\xi(t)$ denotes the variation evolution of the original unscaled 
near-identical system. 
Now the first term of both Eq.~(\ref{solinhomo}) and Eq.~(\ref{solinhomo2}) goes to zero exponentially according to Eq.~(\ref{bdphi}), so that asymptotically we have $\tilde{\xi}(t)=c\xi(t)$, i.e., the variation is scaled by the same factor correspondingly.

\begin{figure}
\includegraphics*[width=0.8\textwidth]{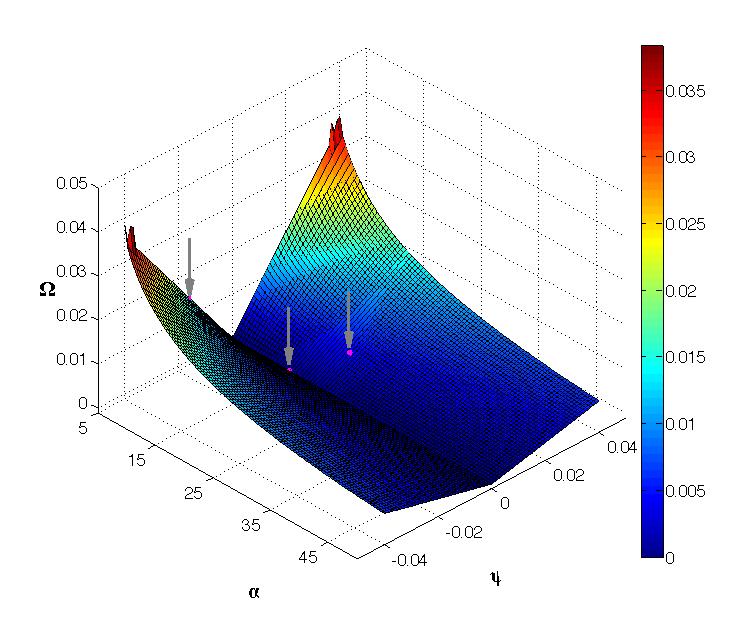}
\caption{
Density plot of extended master stability function $\Omega(\alpha,\psi)$ associated with arbitrary networks of near-identical Lorenz systems.  It is estimated by $\sqrt{\frac{1}{T}\int_{0}^{T}{||\xi(t)||^2dt}}$ with $T=200$ ($||.||$ denotes the Euclidean norm), where $\xi(t)$ is obtained by numerically integrating Eq.~(\ref{extmsf}) with a time step of $0.001$ and discarding initial transient.  Here we have used the coupling function $H(x) = x$. 
}
\label{lorenzmsf}
\end{figure}

\begin{figure}
\includegraphics*[width=0.65\textwidth]{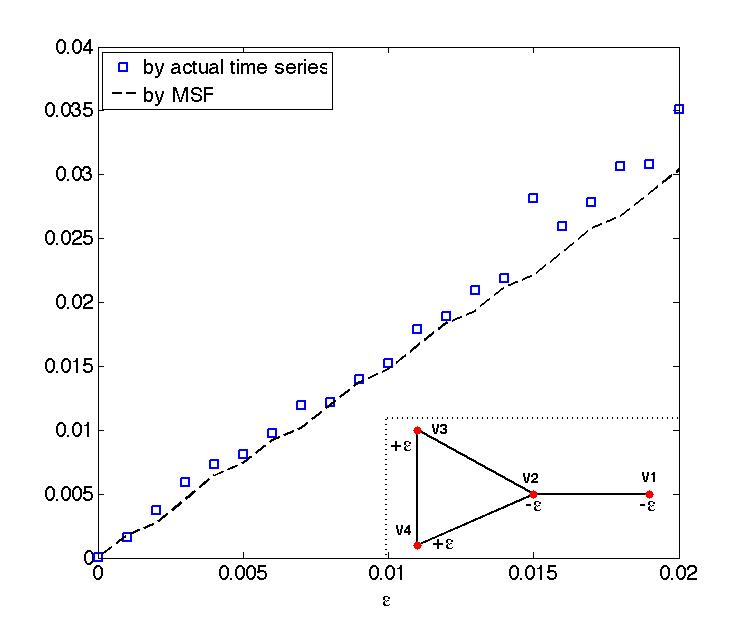}
\caption{
Comparison of predicted synchronization error with actual error.  
For the 4-node network shown in the inset, the error prediction $\sum_{i=2}^4 \Omega(\alpha_i, \psi_i)^2$ (dashed line) was computed using $\Omega$ displayed in Fig.~\ref{lorenzmsf}.
The values of $\alpha_i$ and $\psi_i$ (shown in Fig.~\ref{lorenzmsf} by arrows for $\varepsilon = 0.01$) were obtained from the Laplacian eigenstructure and the parameter mismatch pattern indicated in the inset as a function of $\varepsilon$.
Actual error (squares) was estimated by $\frac{1}{T} \int_0^T \sum_{i=1}^4 || \eta_i (t) ||^2 dt$ with $T=200$ computed from numerical integration of the full system~(\ref{origdyn}) after discarding initial transient.
We used $g=5$ in all calculations.
}
\label{lorenznet}
\end{figure}

{\it Examples of Application ---}
We consider each individual unit $w=[x,y,z]^T$ governed by the Lorenz equations:
\begin{eqnarray}\label{lorenz}
	\dot{x} &=& \sigma(y-x), \nonumber\\
	\dot{y} &=& x(r-z)-y, \nonumber\\
	\dot{z} &=& xy-\beta z,
\end{eqnarray}
where parameters $\sigma=10$, $\beta=\frac{8}{3}$, and 
we consider mismatch between 
units in $r$,
i.e., $r$ corresponds to $\mu$ in Eq.~(\ref{origdyn}). So we have
\begin{eqnarray}
	D_{w}f = 
		\left[ \begin{array}{ccc}
		-\sigma & \sigma & 0 \\
		r-z & -1 & -x \\
		y & x & -\beta \end{array} \right]
\end{eqnarray}
and $D_{\mu}f = [0,x,0]^T$. The coupling function $H$ is taken 
to be $H(x)=x$, 
so that $DH(s)=I_3$ ($\forall s$).
With these choices of $f$ and $H$, we numerically integrate Eq.~(\ref{extmsf}) for a range of $\alpha$ and $\psi$ and estimate the asymptotic norm of $\xi(t)$, which gives 
$\Omega(\alpha,\psi)$ shown in Fig.~\ref{lorenzmsf}.
As shown in Fig.~\ref{lorenznet} for a 4-node network example in the inset, this estimated $\Omega(\alpha,\psi)$, combined with Eq.~(\ref{err}) gives fairly good 
predictions for the actual synchronization error
in the full system~(\ref{origdyn}).
In addition, Fig.~\ref{lorenznet} confirms that the actual synchronization error scales linearly with the magnitude of the parameter mismatch, as predicted by our analysis.

{\it Summary and Discussion ---}
In this letter we have analyzed the stability of synchronization 
in a network of coupled near-identical dynamical systems. 
We have shown that the well-known master stability approach can be extended to this general 
case, allowing us to solve the part of the problem that depends on the individual node dynamics, independently of the network structure and the parameter mismatch pattern over the network.
We have demonstrated the validity of our analysis using a small example network of coupled Lorenz systems.
The extended MSF gives simplified, accurate, and practical estimate of the magnitude of variation 
in a near-identical 
system, provided that the corresponding identical system undergoes stable synchronization according to the original MSF analysis. 
Furthermore, 
our results highlight
the relevance of 
the Laplacian eigenvector structure, in addition to the full eigenvalue spectrum,
in determining the amount of dynamical variation due to parameter mismatch among individual  
dynamics.  This suggests that detailed knowledge of the graph structure may be important
for the design of robust and reliable systems.

{\it Acknowledgements ---}
J.S. and E.M.B have been supported by the Army Research Office 51950-MA. We thank Joseph D. Skufca and Scott R. Fulton for discussion.
%

\end{document}